\documentclass[aps,prl,twocolumn,superscriptaddress]{revtex4-1}
\usepackage[colorlinks,pdfusetitle,urlcolor=blue,citecolor=blue,linkcolor=blue,plainpages=false]{hyperref}
\usepackage{amssymb,graphicx,braket,amsmath}

\begin{document}

\title{Open quantum systems decay across time}
\author{Juliane Klatt}
\thanks{The author recently moved to the Department for Biosystems Science and Engineering, ETH Z\"urich, Mattenstr. 26, CH-4058, Basel, Switzerland}
\affiliation{Physikalisches Institut, Universit\"{a}t Freiburg, Hermann-Herder-Str. 4, D-79104, Freiburg, Germany}
\author{Chahan M. Kropf}
\affiliation{Istituto Nazionale di Fisica Nucleare, Sezione di Pavia, via Bassi 6, I-27100 Pavia, Italy}
\affiliation{Department of Physics, Universit\`a Cattolica del Sacro Cuore, Brescia I-25121, Italy}
\author{Stefan Y. Buhmann}
\affiliation{Physikalisches Institut, Universit\"{a}t Freiburg, Hermann-Herder-Str. 4, D-79104, Freiburg, Germany}
\date{\today}

\begin{abstract}
The description of an open quantum system's decay almost always requires several approximations as to remain tractable. Here, we first revisit the meaning, domain and seeming contradictions of a few of the most widely used of such approximations: semi-group Markovianity, linear response theory, Wigner--Weisskopf and rotating-wave approximation. Secondly, we derive an effective time-dependent decay theory and corresponding generalized quantum regression relations for an open quantum system linearly coupled to an environment. This theory covers all timescales, and subsumes the Markovian and linear-response results as limiting cases. Finally, we apply our theory to the phenomenon of quantum friction.
\end{abstract}

\maketitle

One of the main challenges in the theory of open quantum systems is to identify and apply the appropriate approximations to describe their dynamics. For instance, the description of dynamics in the large-time limit can be tackled by means of a semi-group Markovian approach, or linear response theory (LRT). However, while the former in combination with the quantum regression theorem (QRT) \cite{Lax1963} yields constant exponential decay of excited states and correlations, the latter eventually predicts algebraic decay. This semblant contradiction, which often is not addressed within the open quantum system community, had led to a long-standing debate in the field of quantum friction, which concerned the precise velocity dependence of the friction force experienced by a ground-state atom moving parallel to a macroscopic body at zero temperature \cite{Scheel2009,Tomassone1997,Farias2016,Intravaia2014,Barton2010,Schaich1981}.

As we illustrate in this manuscript, the aforementioned contradiction is a spurious one. It solely arises if results of either the Markov approximation or LRT are extended beyond the respective temporal regimes which they are bound to. While it is known that both semi-group Markovianity and LRT presuppose a large-time limit \cite{Breuer2002}, little attention is paid to the fact that the two limits -- the one referred to in Markovian contexts and the other in LRT -- do not coincide. To complicate things even further, this abuse of terminology often results in a confusion concerning the Markov approximation and its relation to both the Wigner--Weisskopf (or ``pole'') approximation (WWA) \cite{Weisskopf1930} and the rotating-wave approximation (RWA) \cite{Breuer2002}.

The aim of this manuscript is hence three-fold. First, we summarize known results to clearly pin-point the origin of the apparent contradiction discussed above. Secondly, we combine the time-convolutionless projection operator technique (TCL) with results from LRT in order to develop an effective theory of a weakly coupled open quantum system's decay, that is valid on all timescales. Remarkably, this theory yields single-operator dynamics as well as the dynamics of two-point correlations via a generalized QRT. Our approach subsumes the aforementioned approximations as well-defined asymptotes, where we derive two explicit transition times demarcating the regimes referred to by the semi-group Markovian large-time limit and the LRT large-time limit, respectively.  Lastly, we relate our formalism to the case of an atom in body-assisted vacuum and the phenomenon of quantum friction as a case in point.

Both the original QRT and LRT constitute means to arrive at a simplified description of two-point correlations within an open quantum system -- which in absence of these simplifications were to remain intractable. Therefore, we begin by scrutinizing such correlation functions of a system $S$ embedded in an environment $E$, described by the Hamiltonian
\begin{equation}
	\hat H=\hat H_S+\hat H_E+\hat H_\text{int},
\end{equation}
where the $\hat H_{S/E}$ are the Hamiltonians of the isolated system and environment, respectively, and $\hat H_\text{int}$ comprises all interactions. All measurements are conducted on $S$. With the full state of system and environment being represented by the density matrix $\rho$, the expectation values of all operators $\hat A$ acting on $S$ are given by $\langle\hat A\rangle=\text{tr}_S\{\hat A\rho_S\}$ where the reduced density matrix $\rho_S$ is defined as $\rho_S=\text{tr}_E\{\rho\}$. In the following, we consider the scenario of linear coupling between system and environment. This includes phenomena such as linear dipole-coupling of an atom and a surrounding electromagnetic field \cite{Cohen2001,Buhmann2013}, a quantum dot and a Fermi-bath \cite{Scheer2017}, or the spin-boson model \cite{Weiss1999}. It excludes, on the other hand, non-linear phenomena such as the Purcell effect \cite{Bloembergen1948}, Kerr effect \cite{Kerr1875}, and similar non-linear optical effects  \cite{Drummond2014}.

The QRT states that the two-point correlation functions of a semi-group Markovian system obey the same dynamical equations as the expectation values of the system's single observables \cite{Lax1963}. These dynamical equations are the Gorini--Kossakowski--Sudarshan--Lindblad quantum master equations for $\rho_S$ under the influence of an environment $E$ \cite{Gorini1976,Lindblad1976}
\begin{equation}\label{eq:Lindblad}
	\dot\rho_S(t)=
	\frac{[\hat H',\rho_S]}{i\hbar}+\sum_{n} \gamma_n\left(\hat A_n\rho_S\hat A^\dag_n-\frac{1}{2}\{\hat A_n^\dag \hat A_n,\rho_S\}\right).
\end{equation}
While the first term in (\ref{eq:Lindblad}) comprises the unitary part of the reduced dynamics, the non-negative rates $\gamma_n$ and their associated Lindblad operators $\hat A_n$ describe the open system's non-unitary channels (such as relaxation in its different decay modes and decoherence). Both the relaxation rates and the Lindblad operators $\hat A_n$ are time-independent and thereby guarantee the semi-group property of the dynamics. Note that according to modern open quantum systems theory it is not sufficient to have time-dependent rates in order to obtain non-Markovian dynamics \cite{Rivas2014,Breuer2016,Li2018}. In this manuscript, however, we will use the term "Markovian" in the strict sense as a synonym for Eq.~(\ref{eq:Lindblad}) and the resulting semi-group dynamics. The expectation value dynamics of Heisenberg operators $\hat A_n$ acting on such a Markovian system $S$ can be written as a closed system of linear first-order differential equations \cite{Breuer2002},
\begin{equation}
	\frac{d}{dt}\langle\hat A_n(t)\rangle=\sum_{n'}G_{nn'}\langle\hat A_{n'}(t)\rangle,
\end{equation}
with some coefficient matrix $G$ and the two-point correlation functions any $\hat A_n$ and $\hat A_{n'}$ satisfy the very same system of differential equations \cite{Breuer2002},
\begin{equation}\label{eq:QRT}
	\frac{d}{d\tau}\langle\hat A_n(t+\tau)\hat A_{n^{\prime\prime}}(t)\rangle=\sum_{n'}G_{nn'}\langle\hat A_{n'}(t+\tau)\hat A_{n^{\prime\prime}}(t)\rangle.
\end{equation}
Since the quantum master Eq.~(\ref{eq:Lindblad}) generates an exponential relaxation of the system $S$ with the non-negative and constant rates $\gamma_n$ on all timescales, the quantum regression theorem implies an equally exponential decay of system correlations.

If we consider the special case of a two-point correlation function of the same observable at different times, $C_A(\tau)=\langle\hat A(\tau)\hat A(0)\rangle$, LRT provides an alternative means to infer its dependence on the time delay $\tau$. LRT relates the correlation function $C_A$ of the observable $A$ to its power density spectrum $S_A$, which in turn is related to the observable's linear response function $\chi_A$ via the fluctuation-dissipation theorem (FDT) \cite{Kubo1966}
\begin{align}\label{eq:CA}
	C_A(\tau)
	&=\int_{-\infty}^\infty d\omega S_A(\omega)e^{-i\omega\tau}\\
	&=\frac{\hbar}{\pi}\int_0^\infty d\omega \text{Im}[\chi_A(\omega)]e^{-i\omega\tau}.
\end{align}
The linear response function describing the open system's reaction to the perturbation posed by its environment is defined by
\begin{equation}
	\chi_A(\omega)\equiv\frac{i}{2\pi\hbar}\int_0^\infty d\tau\langle[\hat A(\tau),\hat A(0)]\rangle e^{i\omega\tau}.
\end{equation}
For real-valued perturbations and hermitian operators $\hat A$, the response function in time domain must be real-valued and retarded. These features translate into the Schwartz-reflection property $\chi^*_A(\omega)=\chi_A(-\omega^*)$ of the response function in the spectral domain. This in turn implies that the power spectral density $S_A$ must be anti-symmetric with respect to $\omega=0$.

\begin{figure}
	\includegraphics[width=.48\textwidth]{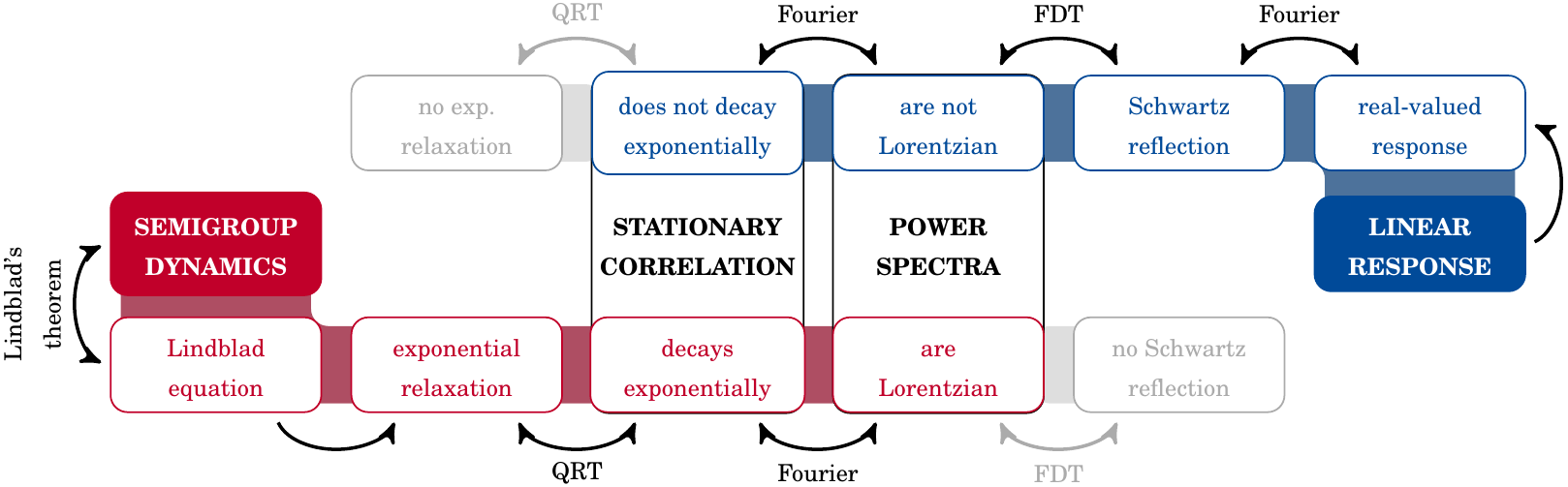}
	\caption{Markovian quantum dynamics are incompatible with LRT. According to the QRT, Markovianity implies exponential decay of correlation functions and hence Lorentzian power spectra. LRT, however, requires non-Lorentzian spectra.}
	\label{fig:QRTFDT}
\end{figure}

It is precisely here, where the Markovian and LRT approaches to obtaining correlation functions clash. Recall that Markovianity and the QRT inevitably arrive at a constant exponential decay of correlations $C_A$. In order to arrive at such a correlation function within LRT and hence by means of Eq.~(\ref{eq:CA}), the underlying spectral density $S_A$ would have to be perfectly Lorentzian. Such $S_A$, however, is not anti-symmetric with respect to $\omega=0$ as required by LRT. Hence, the exponential decay stemming from the Markov approximation and QRT is fundamentally incompatible with the concept of a response function in the sense of LRT (see Fig.~\ref{fig:QRTFDT}). 
The incompatibility of the QRT and LRT approaches arises from the fact that the underlying approximations adhere to different temporal regimes. This is well known from, e.g., atomic spontaneous emission theory. There, time-dependent perturbation theory predicts a Gaussian relaxation yet is restricted to timescales much smaller than the atom's lifetime \cite{Breuer2002}. Markovian quantum master equations, instead, lead to constant exponential decay yet are restricted to times greater than the atom’s lifetime. And eventually, on very large timescales, Markovian dynamics fail again as exponential relaxation gives way to power-law decay, predicted by emission spectra obtained from LRT \cite{Khalfin1958,Knight1976}. The three methods rely on mutually exclusive assumptions and must not be applied beyond the temporal regime they are bound to. In a more spectral than temporal language this problem has already been discussed in references \cite{Talkner1986,Ford1996,Lax2000}, where it translates to the notion of a near-resonance condition implied by the use of the Markov approximation and the quantum regression theorem.

We now reconcile the QRT and LRT approaches to obtaining correlations in open quantum systems, by (i) presenting a formalism for heuristically constructing time-dependent decay rates and level shifts describing expectation value dynamics of open quantum system observables across all timescales, and (ii) deriving a generalized QRT relating these dynamics to two-point correlation functions of these observables. For the sake of simplicity, we focus on a two-level system with level spacing $\hbar\omega_S$ and express the coupling to the environment in rotating-wave approximation (RWA),
\begin{equation}\label{eq:Hint}
\hat H_\text{int}=-\int_0^\infty d\omega[d^*\hat \sigma_+b(\omega)\hat a(\omega)+d\hat \sigma_-b^*(\omega)\hat a^\dag(\omega)]
\end{equation}
and in terms of the dipole moment $\hat D\equiv d^*\sigma_++d\sigma_-$, associated with the two-level system's raising and lowering operators $\hat\sigma_\pm$, and the environment's conjugated observable, $\hat B=\int_0^\infty d\omega[b(\omega)\hat a(\omega)+b^*(\omega)\hat a^\dag(\omega)]$. This implies individual Hamiltonians $\hat H_S=\hbar\omega_S\hat\sigma_+\hat\sigma_-$ and $\hat H_E=\hbar\int_0^\infty d\omega\hat a^\dag(\omega)\hat a(\omega)$. Instead of the Markovian master equation~(\ref{eq:Lindblad}), we begin with a time-dependent one describing the open system's spontaneous decay,
\begin{align}\nonumber
	\frac{d}{d t}\rho_S(t)
	&= i\delta\omega(t)[\hat \sigma_+\hat \sigma_-,\rho_S(t)]\\\label{eq:me}
	&+\Gamma(t)[\hat \sigma_-\rho_S(t)\hat \sigma_+-\frac{1}{2}\{\hat \sigma_+\hat \sigma_-,\rho_S(t)\}],
\end{align}
where $\delta\omega$ and $\Gamma$ are the level shift and decay rate of the system's excited state. Note that generic open quantum systems dynamics include additional pure dephasing terms which we here neglect for simplicity.

By means of TCL, one may expand the shift and rate in Eq.~(\ref{eq:me}) in ordered cumulants (oc) of the system's memory kernel $k(\tau)$ \cite{VanKampen1974a,VanKampen1974b,Breuer2001,Breuer2007,Smirne2010},
\begin{align}\nonumber
	&i\delta\omega_\textrm{tcl}(t)+\Gamma_\textrm{tcl}(t)=\sum_{n=1}^N(-1)^{n+1}\int_0^tdt_1...\int_0^{t_{2n-1}} dt_{2n-1}\\\label{eq:tcl}
	&\qquad\qquad\qquad\times\langle k(t-t_1)...k(t_{2n-2}-t_{2n-1})\rangle_\text{oc}\\
	&k(\tau) = \frac{|d|^2}{\hbar^2}C_B(\tau)e^{i\omega_S\tau}=\frac{|d|^2}{\hbar^2}\int_{-\infty}^\infty d\omega S_B(\omega)e^{-i(\omega-\omega_S)\tau}.
\end{align}
Above, $C_B(\tau)$ and $S_B(\omega)$ denote the auto-correlation function and spectral density of $\hat B$, respectively.
In general, the full TCL expansion (\ref{eq:tcl}) is not known. The series is then cut off and higher-order contributions are neglected. Consecutive orders TCL differ by two time integrals and one memory kernel in the integrand. The expansion parameter is hence given by the kernel's magnitude multiplied with the square of its width. The former is generally proportional to the open system's stationary line width $\Gamma$ while the latter is given by the rate $\gamma$ describing the decay of $C_B$. The expansion parameter of Eq.~(\ref{eq:tcl}) is hence proportional to $\Gamma/\gamma$. Thus, truncating the expansion is only justified in the case of timescale separation---as equally required for the Markov approximation. Yet in stark contrast to Markovian dynamics, Eq.~(\ref{eq:tcl}) in combination with the approximate shifts and rates, $\delta\omega^{(n)}_\text{tcl}$ and $\Gamma^{(n)}_\text{tcl}$, obtained from $n$th-order TCL, provide access to the transient dynamics of the open system.

In the case of time-scale separation, $\delta\omega^{(n)}_\text{tcl}$ and $\Gamma^{(n)}_\text{tcl}$ will be asymptotically constant. While capable of capturing non-exponential decay during the open system's transient dynamics, as well as the stationary---indeed Markovian---behavior of the open system after the completion of its transients, they will fail to reproduce the asymptotic transition to algebraic decay. This can be remedied with insights from LRT. Once stationarity is reached, it is meaningful to speak of a spectral density of the open system. A spectral line has formed whose central frequency and width are approximately determined by the asymptotic TCL shifts and rates,
\begin{equation}\label{eq:statRateShifts}
	\delta\omega_\textrm{stat} = \lim_{t\to\infty}\delta\omega^{(n)}_\text{tcl}(t)  ;  \Gamma_\textrm{stat} = \lim_{t\to\infty}\Gamma^{(n)}_\text{tcl}(t).
\end{equation}

The shape of a stationary quantum mechanical two-level system’s spectral density comprises the full information on the system's spontaneous decay after transients. This is reflected in the relation
\begin{equation}\label{eq:c1dyn}
	c_1(t)=\frac{1}{|d|^2}\int_{-\infty}^\infty d\omega S_D(\omega) e^{-i\omega t},
\end{equation}
linking the interaction-picture amplitude $c_1$ of the system's excited state to the spectral density $S_D$ of its dipole operator (see Eq. (2.4) in reference \cite{Khalfin1958}). The (non-exponential) dynamics as prescribed by LRT can be cast into exponential form by employing a time-dependent frequency and rate, $\dot c_1(t)=c_1(0)\exp\{-[i\omega_\text{lrt}(t)+\Gamma_\text{lrt}(t)/2]\}$, with
\begin{equation}\label{eq:lrt}
	i\omega_\text{lrt}(t)+\frac{\Gamma_\text{lrt}(t)}{2}=-\frac{\dot c_1(t)}{c_1(t)}=\frac{i\int d\omega\omega S_D(\omega)e^{-i\omega t}}{\int d\omega S_D(\omega)e^{-i\omega t}}.
\end{equation}
In an experimental setting, $S_D$ can be determined by means of polarizability measurements. Here, we employ one of the simplest line shapes satisfying the LRT requirements---a Drude--Lorentz peak,
\begin{equation}
S_D(\omega)\propto\frac{2\theta(\omega)}{\pi}\frac{|d|^2\Gamma\tilde\omega_S\omega}{(\tilde\omega_S^2+\frac{\Gamma^2}{4}-\omega^2)^2+\Gamma^2\omega^2},
\end{equation}
with shifted frequency $\tilde\omega_S=\omega_S+\delta\omega$. With that, the Fourier integral (\ref{eq:c1dyn}) can be performed,
\begin{align}\label{eq:c1dyn2}
	c_1(t)
	&\propto e^{-(i\tilde\omega_S+\Gamma/2)t}\\\nonumber
	&-\frac{2}{\pi}\int_0^\infty d\xi\frac{\tilde\omega_S\Gamma\xi e^{-\xi t}}{\left(\tilde\omega_S^2+\Gamma^2/4+\xi^2\right)^2-\Gamma^2\xi^2},
\end{align}
and in addition to the exponential pole-term which dominates for small and intermediate $t$, one obtains a non-resonant integral from the imaginary-frequency axis which cannot be solved analytically yet decays as $t^{-2}$ and takes over for large times.

Hence, while truncated TCL dynamics end stationary, LRT dynamics begin stationary. As there cannot exist any discontinuity in the open system's evolution, we can (i) insert $\delta\omega_\text{stat}$ and $\Gamma_\text{stat}$ into Eq.~(\ref{eq:c1dyn2}) to obtain $\delta\omega_\text{lrt}$ and $\Gamma_\text{lrt}$ from Eq.~(\ref{eq:lrt}), and (i) match the time-dependent TCL and LRT level shifts and decay rates, thus forming a heuristic product shift and rate,
\begin{equation}\label{eq:rec}
	\delta\omega_\text{pro}(t)=\frac{\delta\omega_\text{tcl}(t)\delta\omega_\text{lrt}(t)}{\delta\omega_\text{stat}},\quad
	\Gamma_\text{pro}(t)=\frac{\Gamma_\text{tcl}(t)\Gamma_\text{lrt}(t)}{\Gamma_\text{stat}},
\end{equation}
applying to all timescales. This constitutes our first main result.

\begin{figure}
	\includegraphics[width=.48\textwidth]{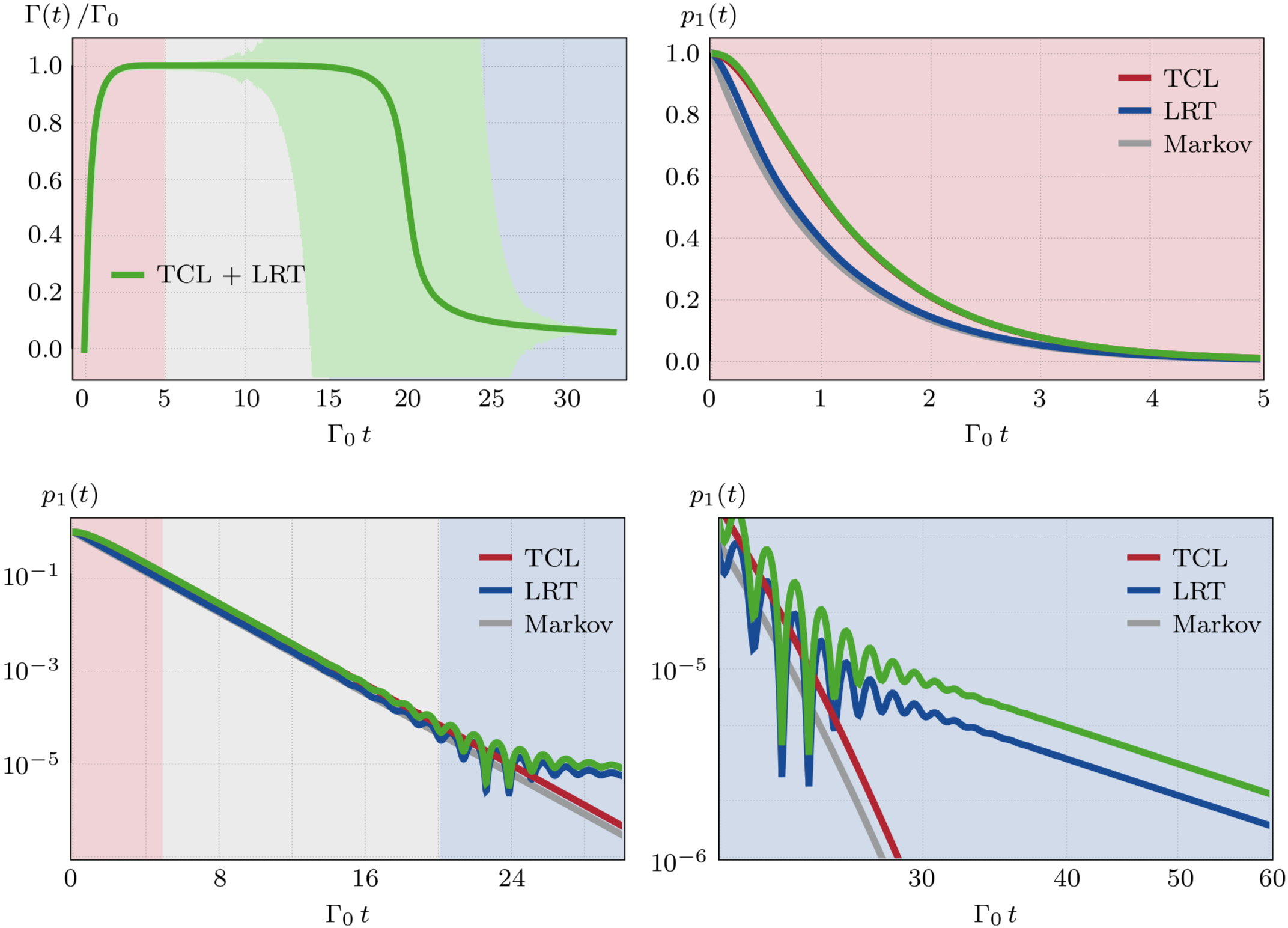}
	\caption{Decay of an initially excited open quantum system according to Markovian, 2$^{nd}$-order TCL, LRT and product rate -- the latter (light green) and its time average (dark) shown in top left -- on short (blue), intermediate (gray) and large (blue) timescales. $S_B$ and $S_D$ are chosen Drude-Lorentzian with zero detuning and $\omega_S+\delta\omega_\text{stat}=100\Gamma_\text{stat}=10\gamma$.}
	\label{fig:decay}
\end{figure}

As an illustration, consider Drude--Lorentzian shapes for both $S_B$ and $S_D$. Fig.~\ref{fig:decay} displays the corresponding product rate for $n=2$ as well as resulting dynamics of the excited state population $p_1=|c_1|^2$ across timescales. As expected, the TCL and product rate capture the Gaussian decay on transient timescales, while the algebraic decay on asymptotic timescales is captured by the LRT and  product rate. The Markovian rate's validity is restricted to the intermediate regime. The difference of the LRT and product rate dynamics on large timescales solely stems from different short-time behavior of the two. The systematic offset in double-logarithmic scale continuously decreases on linear scale, thus ensuring that asymptotically both dynamics coincide -- as they must.
The Markovian large-time limit is reached when the transient have decayed at $t_1 = \Gamma_\textrm{stat}^{-1}$, while the LRT large-time limit only kicks in at $t_2$ when the non-resonant term in Eq.~(\ref{eq:c1dyn2}) takes over at $t_2$ satisfying
\begin{align}
	e^{-\Gamma_\text{stat}t_2/2}=\frac{2(\omega_S+\delta\omega_\text{stat})\Gamma_\text{stat}}{\pi\left[(\omega_S+\delta\omega_\text{stat})^2+\frac{1}{4}\Gamma^2_\text{stat}\right]^2t_2^2}. 
\end{align} These two times $t_1$ and $t_2$ differ drastically.

In order to infer from the time-dependent product shifts and rates not only population yet also correlation dynamics, one requires a generalized QRT. The derivation can be found in the Appendix and results in
\begin{align}
&\frac{\frac{d}{d\tau}\langle\hat\sigma_\pm(t+\tau)\hat\sigma_\mp(t)\rangle}{\langle\hat\sigma_\pm(t+\tau)\hat\sigma_\mp(t)\rangle}\simeq\pm\left[i\omega_\text{S}+i\delta\omega(\tau)\mp\frac{1}{2}\Gamma(\tau)\right],\\
&\frac{\frac{d}{d\tau}\langle\hat\sigma_\pm(t+\tau)\hat\sigma_\pm(t)\rangle}{\langle\hat\sigma_\pm(t+\tau)\hat\sigma_\pm(t)\rangle}\simeq\pm\left[i\omega_\text{S}-i\delta\omega(\tau)\mp\frac{1}{2}\Gamma(\tau)\right],
\end{align}
where the left and right hand sides differ by terms of order $d^4$. This is our second main result. For a ground-state system---as e.g., discussed in the quantum friction debate---the above yields
\begin{equation}
C_D(\tau) \equiv \langle \hat D(0)\hat D(\tau)\rangle = |d|^2e^{-\int_0^\tau\left[i\omega_\text{S}+i\delta\omega(\tau)+\frac{1}{2}\Gamma(\tau)\right]},
\end{equation}
which upon insertion of $\delta\omega_\text{pro}(\tau)$ and $\Gamma_\text{pro}(\tau)$ decays in a Gaussian manner up until $t_1$, then enters a Markovian period which lasts until $t_2$, and finally reaches the algebraic decay regime.

Before we conclude, we contrast the RWA, WWA, and Markov approximation. All approximate a physical system by omitting non-resonant effects. The RWA neglects non-resonant terms in the $\hat{H}_\text{int}$ which would otherwise contain products $\hat\sigma_+\hat a^\dag$ and $\hat\sigma_-\hat a$ allowing for the spontaneous creation and annihilation of pairs of excitations and thereby enabling a ground-state system to fluctuate into its excited state. The Markov approximation additionally imposes a Lorentzian spectral density onto the open quantum system---thereby neglecting non-resonant contributions to the system's excited state dynamics \eqref{eq:c1dyn2}---as well as a decay rate which is only correct up to 2$^{nd}$ order in coupling strength. Hence, while the Markov approximation includes RWA, applying RWA does not imply restricting to Markovian dynamics. The WWA, lastly, is similar to the Markov approximation in that it only keeps the resonant contributions in Eq.~\eqref{eq:c1dyn2}. However, it neither implies RWA dynamics nor a 2$^{nd}$-order truncation of the syetm's decay rate. I.e., if the correct rate were known, applying the WWA would lead to semi-group dynamics yet differ from semi-group Markovian dynamics in the 4$^\mathrm{th}$ order in coupling.

Summarizing, in order to describe the dynamics of an open quantum system linearly coupled to an environment, we have derived effective time-dependent level shifts and decay rates valid on all time scales. We have demonstrated how the large-time limits implied by Markovian and LR theory differ and how ignoring that difference leads to seemingly contradictory results. Lastly, we have derived a generalized QRT for two-point correlation functions. Combining our two main results---the aforementioned effective shifts and rates and the generalized QRT---allows us to report the final resolution of the quantum friction debate. Our effective theory enables us to study quantum friction in settings other than the peculiar case of the (LRT) large-time limit of a ground-state atom moving parallel to a macroscopic body, and thereby shows that the quantum friction force is linear in relative velocity if any of these conditions is lifted. I.e., it is linear if the atom is either excited, and/or moving in a non-parallel manner, and/or has not yet reached the large-time limit in the LRT sense. This is in agreement with both the Markov theory \cite{Scheel2009} and the prediction made with LRT \cite{Intravaia2014}, if correct temporal regimes are taken into consideration, and therefore reconciles the two.

The authors would like to thank D.A.R. Dalvit, F. Intravaia, and H.-P. Breuer, the valuable discussion with whom greatly benefited this work.

\bibliographystyle{}
\bibliography{Open_Quantum_Systems_Decay_across_Time_Notes_Final.bib}

\end{document}


\onecolumngrid
\section{Supplemental Material}
Here we report the keys steps of derivation of the generalized QRT. First, we note that
\begin{align}\nonumber
\langle\hat\sigma^\textrm{heis}_+(t+\tau)\hat\sigma^\textrm{heis}_-(t)\rangle=\textrm{tr}\{U_\textrm{int}^\dag(t+\tau,0)U_0^\dag(t+\tau,0)\hat\sigma_+U_0(t+\tau,t)U_\textrm{int}(t+\tau,t)\hat\sigma_-U_0(t,0)U_\textrm{int}(t,0)\rho\}\\\nonumber
=\textrm{tr}\{\hat\sigma_+^\textrm{int}(t+\tau)U_\textrm{int}(t+\tau,t)[\hat\sigma_-^\textrm{int}(t)\rho^\textrm{int}(t)]U^\dag_\textrm{int}(t+\tau,t)\}+\mathcal{O}(d^4)\\\label{eq:sisibar}
\equiv\textrm{tr}\{\hat\sigma_+^\textrm{int}(t+\tau)U_{\textrm{int},t}(\tau,0)\bar\rho_t(0)U^\dag_{\textrm{int},t}(\tau,0)\}
=\textrm{tr}\{\hat\sigma_+(t+\tau)\bar\rho_t(\tau)\},
\end{align}
where $U_0$ and $U_\textrm{int}$ are defined as the standard interaction-picture evolution operators and the pseudo-density operator $\bar\rho_t(\tau)\equiv\hat\sigma_-(t+\tau)\rho(t+\tau)$ evolves exactly as a true density matrix in the interaction picture. Note that in general the initial condition $\bar\rho(0)$ does not factorize. Taking the derivative of Eq.~\ref{eq:sisibar} with respect to the time delay $\tau$ reveals
\begin{equation}
\frac{d}{d\tau}\langle\hat\sigma_+(t+\tau)\hat\sigma_-(t)\rangle
=\left[i\omega_\textrm{A}+i\delta\omega(\tau)-\frac{1}{2}\Gamma(\tau)\right]\langle\hat\sigma_+(t+\tau)\hat\sigma_-(t)\rangle,
\end{equation}
and analogously
\begin{align}\label{eq:C-+}
\frac{d}{d\tau}\langle\hat\sigma_-(t+\tau)\hat\sigma_+(t)\rangle
=-\left[i\omega_\textrm{A}+i\delta\omega(\tau)+\frac{1}{2}\Gamma(\tau)\right]\langle\hat\sigma_-(t+\tau)\hat\sigma_+(t)\rangle,\\\label{eq:C++}
\frac{d}{d\tau}\langle\hat\sigma_+(t+\tau)\hat\sigma_+(t)\rangle
=\left[i\omega_\textrm{A}-i\delta\omega(\tau)-\frac{1}{2}\Gamma(\tau)\right]\langle\hat\sigma_+(t+\tau)\hat\sigma_+(t)\rangle,\\\label{eq:C--}
\frac{d}{d\tau}\langle\hat\sigma_-(t+\tau)\hat\sigma_-(t)\rangle
=-\left[i\omega_\textrm{A}-i\delta\omega(\tau)+\frac{1}{2}\Gamma(\tau)\right]\langle\hat\sigma_-(t+\tau)\hat\sigma_-(t)\rangle.
\end{align}